\documentclass[%
 aip,
 jap,
 onecolumn,
 amsfonts,
 amsmath,
 amssymb,
 superscriptaddress
 ]{revtex4-1}%

\usepackage{graphicx}
\usepackage{subfigure}
\usepackage{dcolumn}
\usepackage{bm}

\usepackage{SIunits}

\begin{document}

\title{Tunable light trapping and absorption enhancement with graphene-based complementary metamaterials}

\author{Shuyuan Xiao}
\affiliation{Wuhan National Laboratory for Optoelectronics, Huazhong University of Science and Technology, Wuhan 430074, People's Republic of China}

\author{Tao Wang}
\email{wangtao@hust.edu.cn}
\affiliation{Wuhan National Laboratory for Optoelectronics, Huazhong University of Science and Technology, Wuhan 430074, People's Republic of China}

\author{Tingting Liu}
\affiliation{School of Electronic Information and Communications, Huazhong University of Science and Technology, Wuhan 430074, People's Republic of China}

\author{Chaobiao Zhou}
\affiliation{Wuhan National Laboratory for Optoelectronics, Huazhong University of Science and Technology, Wuhan 430074, People's Republic of China}

\author{Xiaoyun Jiang}
\affiliation{Wuhan National Laboratory for Optoelectronics, Huazhong University of Science and Technology, Wuhan 430074, People's Republic of China}

\author{Xicheng Yan}
\affiliation{Wuhan National Laboratory for Optoelectronics, Huazhong University of Science and Technology, Wuhan 430074, People's Republic of China}

\author{Le Cheng}
\affiliation{Wuhan National Laboratory for Optoelectronics, Huazhong University of Science and Technology, Wuhan 430074, People's Republic of China}

\author{Chen Xu}
\affiliation{Department of Physics, New Mexico State University, Las Cruces 88001, United State of America}

\date{\today}

\begin{abstract}
Surface plasmon resonance (SPR) has been intensively investigated and widely exploited to trap the incident light and enhance absorption in the optoelectronic devices. The availability of graphene as a plasmonic material with strong half-metallicity and continuously tunable surface conductivity makes it promising to dynamically modulate the absorption enhancement with graphene-based metamaterials. Here we numerically demonstrate tunable light trapping and absorption enhancement can be realized with graphene-based complementary metamaterials. Furthermore, we also explore the polarization sensitivity in the proposed device, in which case either TM or TE plane wave at the specific wavelength can be efficiently absorbed by simply manipulating the Fermi energy of graphene. Therefore, this work can find potential applications in the next generation of photodetectors with tunable spectral and polarization selectivity in the mid-infrared and terahertz (THz) regimes.
\end{abstract}

\maketitle

\section{Introduction}\label{sec1}
Surface plasmon resonance (SPR), the collective electronic excitation at metal/dielectric interface, provides an effective route to manipulate light-matter interaction.\cite{schaadt2005enhanced,choi2013versatile,ye2017large} The unprecedented ability of SPR to trap the incident light in the near field and induce the effects of electromagnetic field enhancement and light energy concentration has been intensively investigated and widely exploited to enhance light absorption in the optoelectronic devices. In the past, a vareity of metal-based plasmonic metamaterials, such as ribbon, disk, ring, cross and other shapes have been presented to integrate with light-absorbing materials to improve their absorption performance in the infrared and terahertz (THz) regimes.\cite{le2009plasmon,liu2010infrared,khardikov2010trapping,wang2015novel,xiong2015ultrabroadband,wang2016simple} Unfortunately, the resonant responses of metal-based plasmonic metamaterials are dependent on the geometric parameters, and therefore the operating wavelength of these hybrid optoelectronic devices are unchangeable once they are fabricated, which severely hinders flexible applications requiring tunable spectral selectivity in practice.

Graphene is a promising two-dimensional (2D) material with exceptional optical and electrical properties and serves as a building block in the field of modern optoelectronics.\cite{novoselov2004electric,bonaccorso2010graphene} In the mid-infrared and THz regimes, graphene exhibits strong half-metallicity when coupling with the incident light and supports SPR for active applications.\cite{grigorenko2012graphene,tamagnone2012analysis,he2016further} Moreover, the resonant responses of graphene-based plasmonic metamaterials can be dynamically modulated by the continuously tunable surface conductivity of graphene with manipulating its Fermi energy, which are considered as serious competitors to their metal-based counterparts.\cite{ju2011graphene,he2014electrically,he2015tunable,lin2015combined,xia2016excitation_surface,linder2016graphene} Very recently, the pioneering works have demonstrated the possibility for light trapping and absorption engineering with high efficiency and tunable spectral selectivity by integrating graphene-based plasmonic metamaterials with bulky or 2D materials.\cite{zhang2015towards,xiao2016tunable} Nevertheless, the graphene resonators in these designs exist in the isolated fashion, which may not be expediently tuned in practice. In addition, the tunable polarization selectivity in this kind of devices also remains to be further explored.

In this work, we numerically demonstrate light trapping and absorption enhancement can be realized with graphene-based complementary metamaterials. The hybrid device consists of a monolayer graphene perforated with a periodic array of nanoholes on the top of the light-absorbing material separated by an insulating spacer. The simulation results show that the excitation of SPR in the monolayer graphene can effectively trap the incident light in the near field and enhance the absorption in the nearby light-absorbing material by more than one order of magnitude. Furthermore, we explore the polarization sensitivity in the proposed device when transforming the shape of complementary resonators from circular nanohole to elliptical nanohole. With manipulating the Fermi energy of graphene, the polarization-sensitive absorption here can be dynamically modulated over a broad spectral regime, which can find potential applications in the next generation of photodetectors with tunable spectral and polarization selectivity in the mid-infrared and THz regimes.\cite{koppens2014photodetectors,zhu2014electrically_controlling,zhu2014electrically_tunable}

\section{The geometric structure and numerical model}\label{sec2}
\begin{figure}[htbp]
\centering
\includegraphics[scale=0.3]{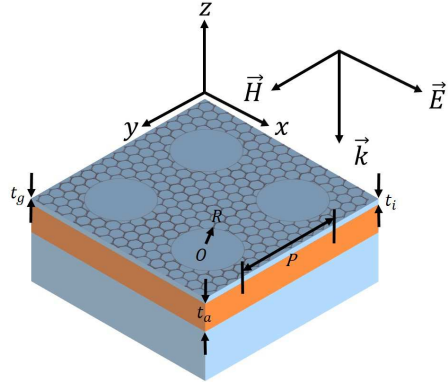}
\caption{\label{fig:1}The schematic geometry of our proposed graphene-based complementary metamaterials. The unit cell consists of a monolayer graphene perforated with a circular nanohole on the top of the light-absorbing material separated by an insulating spacer.}
\end{figure}

FIG.~\ref{fig:1} schematically depicts our proposed graphene-based complementary metamaterials. In this hybrid device, the unit cell is arranged in a periodical array with a lattice constant $P=400$ nm and consists of a monolayer graphene perforated with a circular nanohole on the top of the light-absorbing material separated by an insulating spacer. The radius of the circular nanohole is $R=120$ nm, and the effective thickness of the monolayer graphene is set as $t_{g}=1$ nm. The thicknesses of the insulating spacer and the light-absorbing material are $t_{i}=20$ nm and $t_{a}=100$ nm, respectively, and the substrate is assumed to be semi-infinite. The insulating spacer and the substrate are treated as lossless dielectrics with a real permittivity of $\varepsilon_{d}=1.96$. The light-absorbing material is modeled using a complex permittivity of $\varepsilon_{a}=\varepsilon^{'}+i\varepsilon^{''}$, where $\varepsilon^{'}=10.9$ and $\varepsilon^{''}$ is related to the absorption coefficient $\alpha=0.05$ $\micro\meter$$^{-1}$ accounting for losses, comparable to the typical material Hg$_{1-x}$Cd$_{x}$Te ternary alloy exploited for photodetection in the mid-infrared and THz regimes.\cite{palik1998handbook,rogalski2005hgcdte}

The surface conductivity of graphene can be described with the random-phase approximation (RPA) in the local limit, including both intraband and interband processes\cite{zhang2014coherent,zhang2016towards,zhao2016tunable}
\begin{equation}
    \begin{split}
      \sigma_{g} &=\sigma_{intra}+\sigma_{inter}=\frac{2e^{2}k_{B}T}{\pi\hbar^{2}}\frac{i}{\omega+i\tau^{-1}}\ln[2\cosh(\frac{E_{F}}{2k_{B}T})]\\
      &+\frac{e^2}{4\hbar}[\frac{1}{2}+\frac{1}{\pi}\arctan(\frac{\hbar\omega-2E_{F}}{2k_{B}T}) \\
      &-\frac{i}{2\pi}\ln\frac{(\hbar\omega+2E_{F})^{2}}{(\hbar\omega-2E_{F})^{2}+4(k_{B}T)^{2}}],\label{eq1}
    \end{split}
\end{equation}
where $e$ is the charge of an electron, $k_B$ is the Boltzmann constant, $T$ is the operation temperature, $\hbar$ is the reduced Planck's constant, $\omega$ is the angular frequency of the incident light, $\tau$ is the carrier relaxation time and $E_F$ is the Fermi energy. In the lower THz regime, the contribution originated from the interband process can be safely neglected due to the Pauli exclusion principle and the surface conductivity is reduced to a Drude-like model,\cite{hanson2008quasi,xiao2017strong}
\begin{equation}
    \sigma_{g}=\frac{e^{2}E_{F}}{\pi\hbar^{2}}\frac{i}{\omega+i\tau^{-1}},\label{eq2}
\end{equation}
where the carrier relaxation time $\tau=(\mu E_{F})/(e v_{F}^{2})$ is dependent on the carrier mobility $\mu=10000$ cm$^{2}$/V$\cdot$s, the Fermi energy $E_{F}$ and the Fermi velocity $v_{F}=10^{6}$ m/s.\cite{zhao2016graphene,huang2017tunable} Furthermore, the anisotropic permittivity of graphene can be described in a diagonal tensor form, with isotropic dispersive components in the plane and non-dispersive component out of the plane\cite{gao2012excitation,zeng2014high,xia2016excitation_crest}
\addtocounter{equation}{1}
\begin{align}
\varepsilon_{xx}=\varepsilon_{yy}&=2.5+\frac{i \sigma_{g}}{\varepsilon_{0}\omega t_{g}}, \tag{\theequation a}\\
\varepsilon_{zz}&=2.5, \tag{\theequation b}\label{eq3}
\end{align}
where $\varepsilon_{0}$ is the permittivity of vacuum.

The finite-difference time-domain method is exploited for full-wave numerical simulations (FDTD Solutions, Lumerical Inc., Canada). In the calculations, the moderate mesh grid is adopted to balance the simulation time and accuracy. The periodical boundary conditions are employed in the $x$ and $y$ directions and perfectly matched layers are utilized in the $z$ direction along the propagation of the incident plane wave.

\section{Simulation results and discussions}\label{sec3}
\begin{figure}[htbp]
\centering
\includegraphics[scale=0.3]{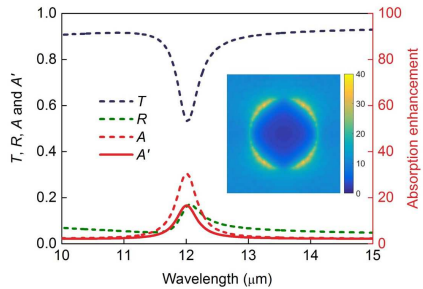}
\caption{\label{fig:2}The simulated spectra of the transmission $T$, the reflection $R$, the total absorption $A$ and the absorption in the light-absorbing material $A^{'}$ with the absorption coefficient $\alpha=0.05$ $\micro\meter$$^{-1}$ and the Fermi energy of graphene $E_F=0.6$ eV. The enhancement factor of the absorption in the light-absorbing material is also shown compared to that in impedance matched media. The inset plots the simulated electric field distribution in the $x$-$y$ plane ($|E_{z}|$) at the resonance.}
\end{figure}

With the electric field of the incident light oriented along the $x$ axis, ($x$-polarized, i.e., transverse-magnetic (TM) plane wave), SPR in the monolayer graphene is excited in the spectral regime of interest. FIG.~\ref{fig:2} illustrates the numerically simulated spectra with an initial Fermi energy of graphene $E_{F}=0.6$ eV. The resonance occurs at around 12 $\micro\meter$, arising a strong transmission suppression and absorption enhancement. The total absorption is $A=31\%$ at the resonance and the absorption in  the light-absorbing material is $A^{'}=16.9\%$. Note that the light-absorbing material here is assumed to be only 100 nm thick with the absorption coefficient $\alpha=0.05$ $\micro\meter$$^{-1}$, corresponding to an extreme weak absorption of about $1\%$ in impedance matched media, an enhancement factor of 16.9 can be achieved at the resonance. Here the absorption enhancement should be attributed to the excitation of SPR in the monolayer graphene. The simulated $x$-$y$ plane electric field distribution ($|E_{z}|$) at the resonance in the inset shows a strong enhancement of ($|E_{z}|$) around the circular nanohole, which is symmetric to the $y$-axis. This is a characteristic behavior of the excitation of SPR in the nanohole-shaped resonator, which results from the accumulated charges around the circular nanohole due to the TM plane wave.\cite{zhang2016towards} Hence, SPR effectively trap the light energy and provide sufficient time to dissipate it by the losses in the light-absorbing material.

\begin{figure}[htbp]
\centering
\includegraphics[scale=0.3]{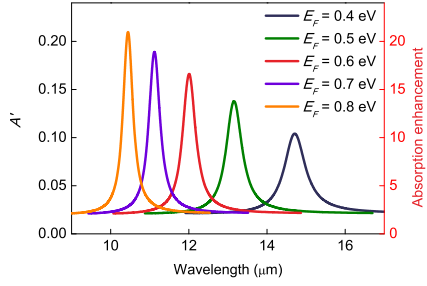}
\caption{\label{fig:3}The simulated spectra of the absorption in the light-absorbing material $A^{'}$ with different Fermi energies of graphene $E_{F}$. The enhancement factor of the absorption in the light-absorbing material is also shown compared to that in impedance matched media.}
\end{figure}

The optical properties of graphene are highly dependent on its Fermi energy, which is the origin of the dynamical tunability of SPR in the graphene-based metamaterials. To demonstrate the spectral tunability of our proposed hybrid device, FIG.~\ref{fig:3} illustrates the absorption enhancement in the light-absorbing material with different Fermi energies of graphene. When $E_{F}$ starts at 0.4 eV, the resonance locates at 14.7 $\micro\meter$ and the absorption in the light-absorbing material is $10.5\%$. As $E_{F}$ increases to 0.7 eV, the resonance blue shifts to 11.1 $\micro\meter$ and the absorption goes up to $19.1\%$. Finally, when $E_{F}$ comes to 0.8 eV, the absorption reaches $21.3\%$ at 10.4 $\micro\meter$ with an enhancement as high as 21 times. It can be seen that with the increase of Fermi energy, the resonance wavelength becomes shorter and absorption enhancement increases simultaneously. The physics mechanism lies in that as the Fermi energy of graphene increases, the wavelength of SPR grows longer (for a fixed frequency or vacuum wavelength) and the nanohole-shaped resonator looks relatively smaller for the incident light, hence the resonance experiences an obvious blue shift. Meanwhile, the increase of Fermi energy also leads to the larger surface conductivity of graphene and SPR becomes less lossy, thus the number of charge carriers contributing to the resonance increases. As a result, the field enhancement with higher $E_{F}$ are stronger than that with lower $E_{F}$, which results in a higher absorption enhancement in the light-absorbing material. Therefore, the tunable light trapping and absorption enhancement can be achieved by manipulating the Fermi energy of graphene, without changing the geometrical parameters of complementary metamaterials. Here, we would like to highlight that the graphene nanohole-shaped resonators exploited in our proposed hybrid device is in the monolayer morphology rather than isolated fashion, which is much easier to fabricate and manipulate than previous studies.

\begin{figure}[htbp]
\centering
\includegraphics[scale=0.3]{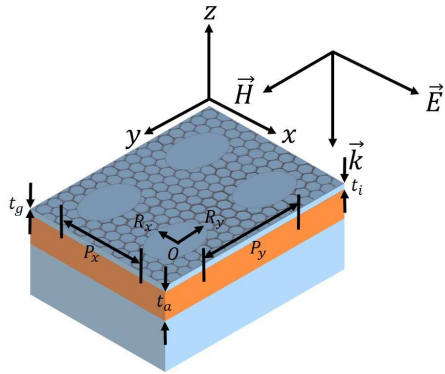}
\caption{\label{fig:4}The schematic geometry of our proposed anisotropic graphene-based complementary metamaterials. In the unit cell, the monolayer graphene is perforated with a elliptical nanohole, and the rest keep the same as previous settings in FIG.~\ref{fig:1}.}
\end{figure}

\begin{figure}[htbp]
\centering
\subfigure[]{ \label{fig:5subfig:1} 
\includegraphics[scale=0.3]{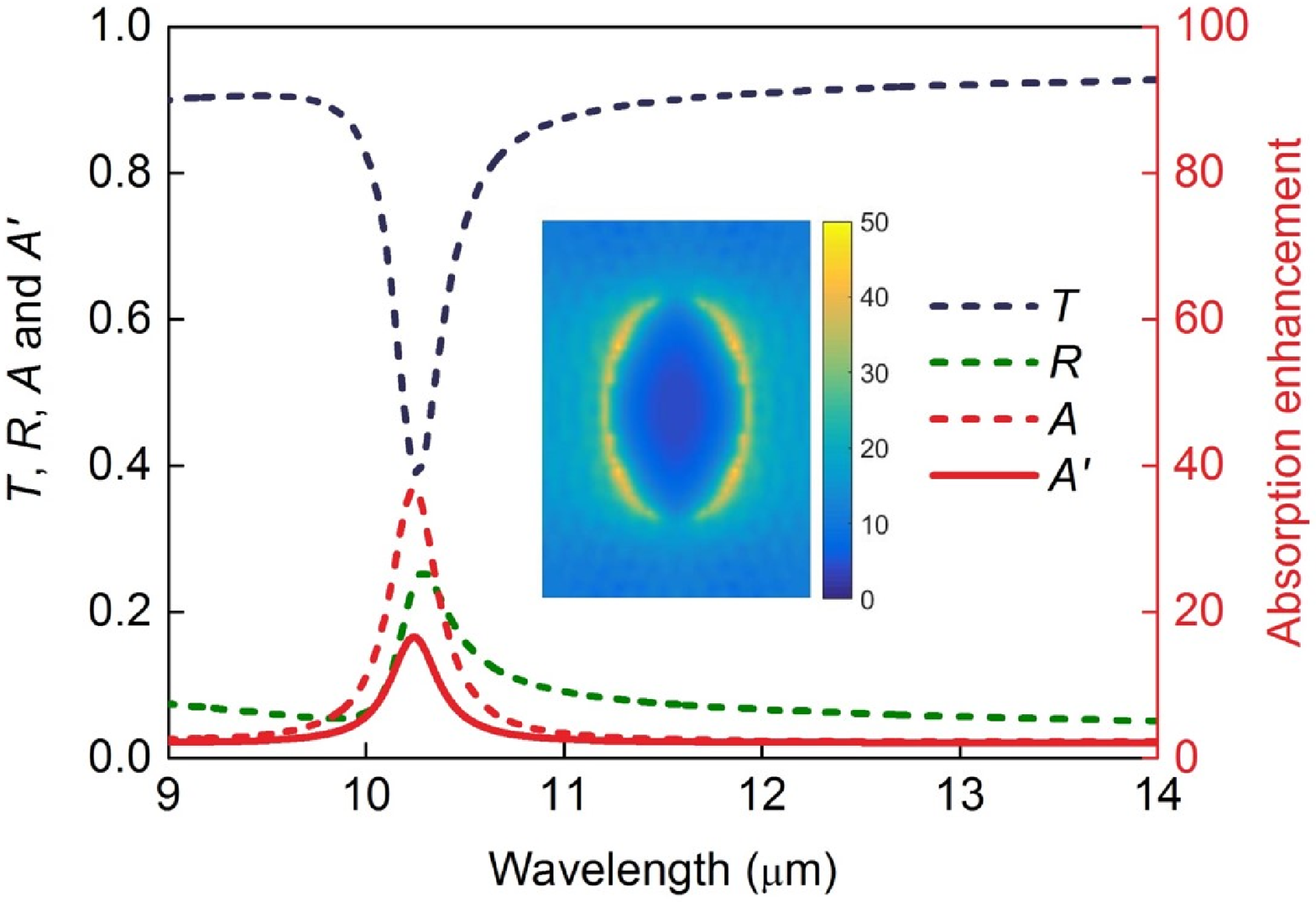}}
\subfigure[]{ \label{fig:5subfig:2} 
\includegraphics[scale=0.3]{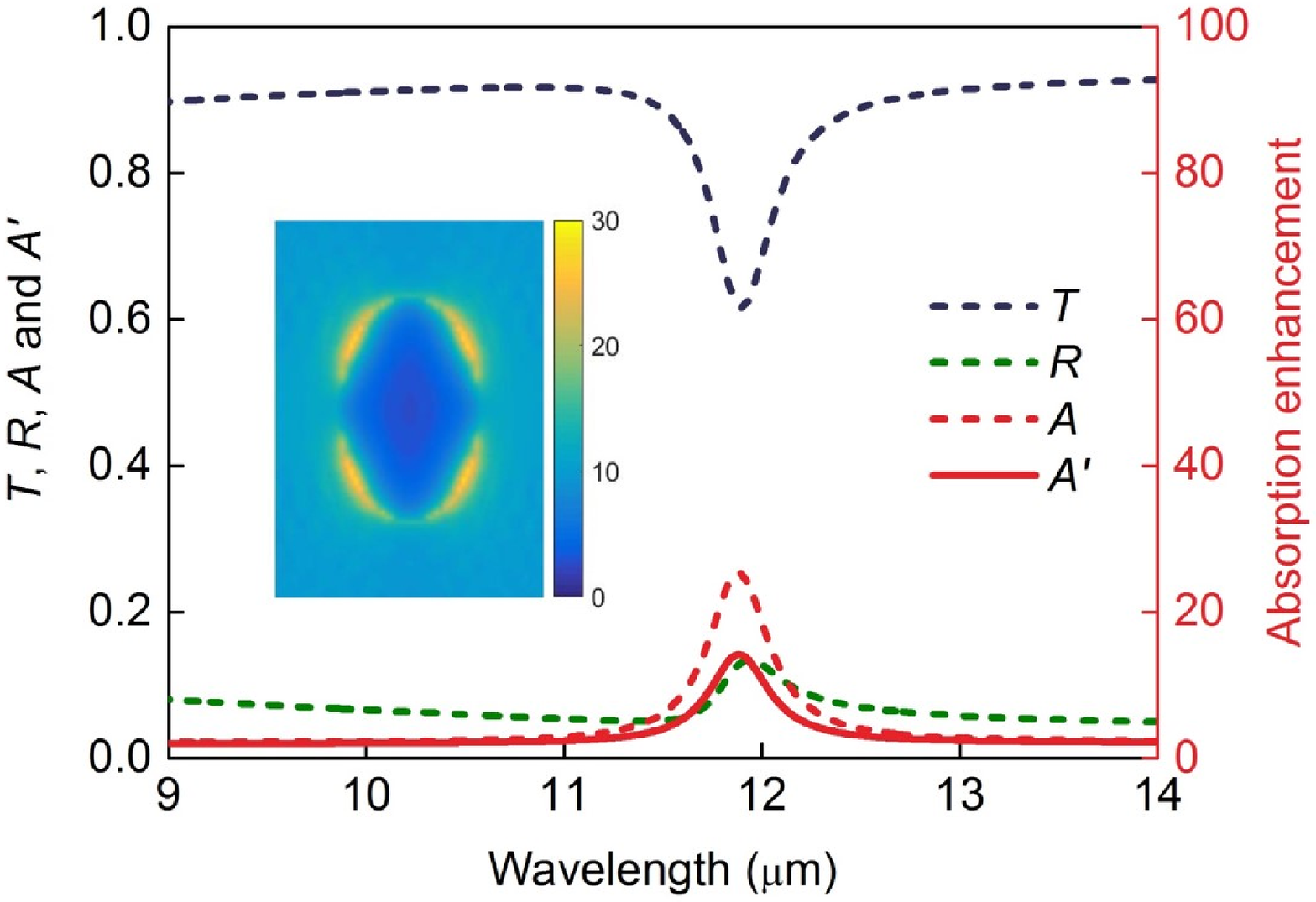}}
\caption{\label{fig:5}The simulated spectra of the transmission $T$, the reflection $R$, the total absorption $A$ and the absorption in the light-absorbing material $A^{'}$ with the absorption coefficient $\alpha=0.05$ $\micro\meter$$^{-1}$ and the Fermi energy of graphene $E_F=0.6$ eV for (a) TM and (b) TE plane wave. The enhancement factor of the absorption in the light-absorbing material is also shown compared to that in impedance matched media. The inset plots the simulated electric field distribution in the $x$-$y$ plane ($|E_{z}|$) at the resonance.}
\end{figure}

To further explore the polarization sensitivity in the proposed device, we transform the shape of complementary resonators from circular nanohole to elliptical nanohole. As shown in FIG.~\ref{fig:4}, the monolayer graphene is perforated with a periodical array of elliptical nanoholes, where the lattice constants are $P_{x}=300$ nm and $P_{y}=400$ nm. The long axis and short axis of the elliptical nanoholes are $R_{x}=80$ nm and $R_{y}=120$ nm accordingly. The thicknesses of the insulating spacer and the light-absorbing material, as well as their optical properties, keep the same as previous settings. The numerically simulated spectra with an initial Fermi energy of graphene $E_{F}=0.6$ eV are depicted in FIG.~\ref{fig:5subfig:1} and \ref{fig:5subfig:2} for TM and TE plane wave at normal incidence, respectively. For TM plane wave, the resonance happens at around 10.2 $\micro\meter$ and the absorption in the light-absorbing material is $16.8\%$. For TE plane wave, the resonance at 11.9 $\micro\meter$ shows the maximum absorption in the light-absorbing material as $14.4\%$. The $x$-$y$ plane electric field distribution ($|E_{z}|$) at the resonances are plotted in the insets, where the plasmon-excited electric field enhancement around the elloptical nanoholes are symmetric to the $y$-axis under TM incident plane wave and symmetric to the $x$-axis under the TE incident plane wave, respectively. Hence, the absorption enhancement in the light-absorbing material for both the TM and TE mode should be ascribed to the excitation of SPR in the monolayer graphene. The different resonance wavelengths for TM and TE modes arise from the different values of the long axis and short axis of the elliptical nanohole-shaped resonator in the $x$ and $y$ directions.

\begin{figure}[htbp]
\centering
\subfigure[]{ \label{fig:6subfig:1} 
\includegraphics[scale=0.3]{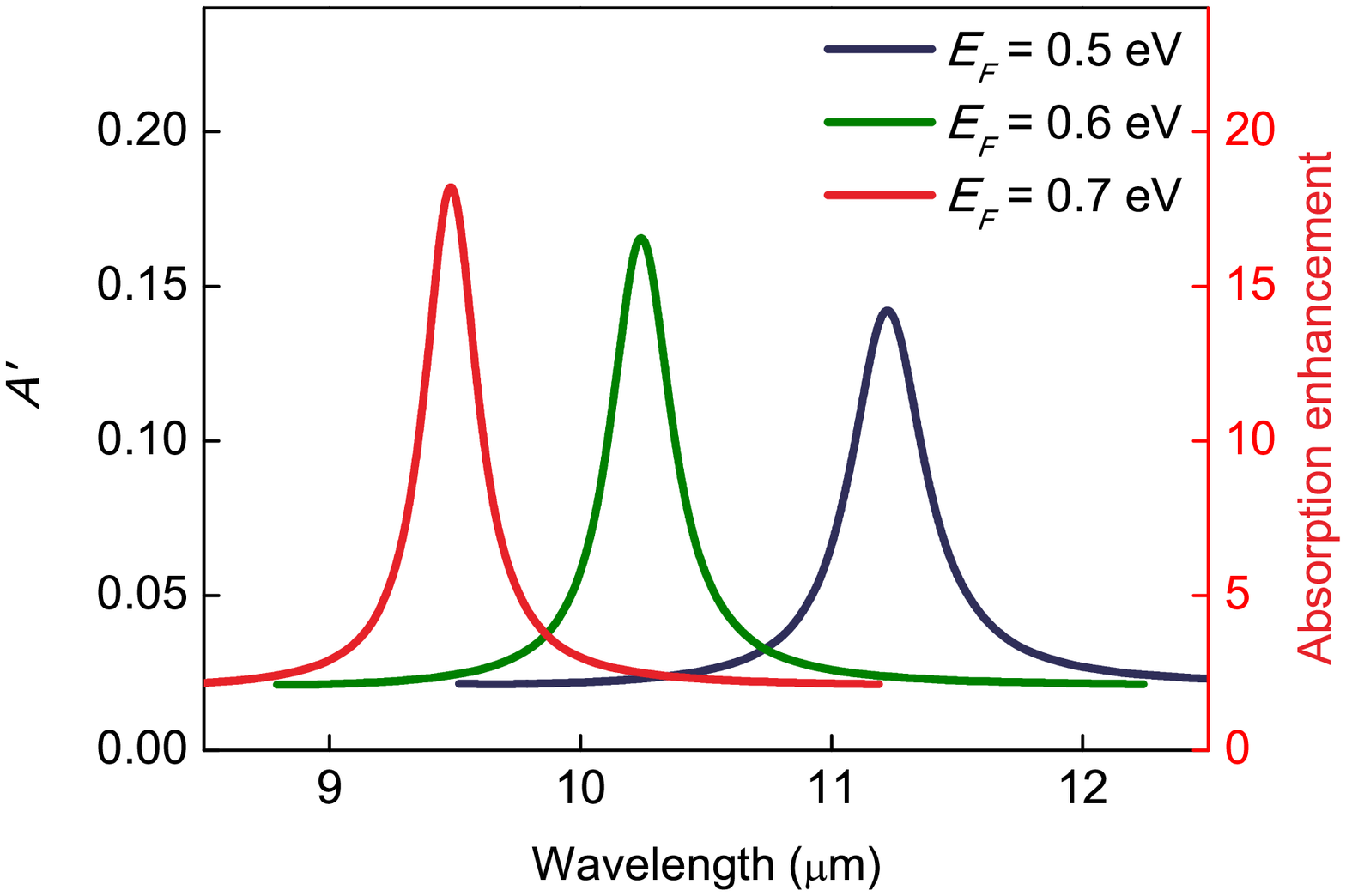}}
\subfigure[]{ \label{fig:6subfig:2} 
\includegraphics[scale=0.3]{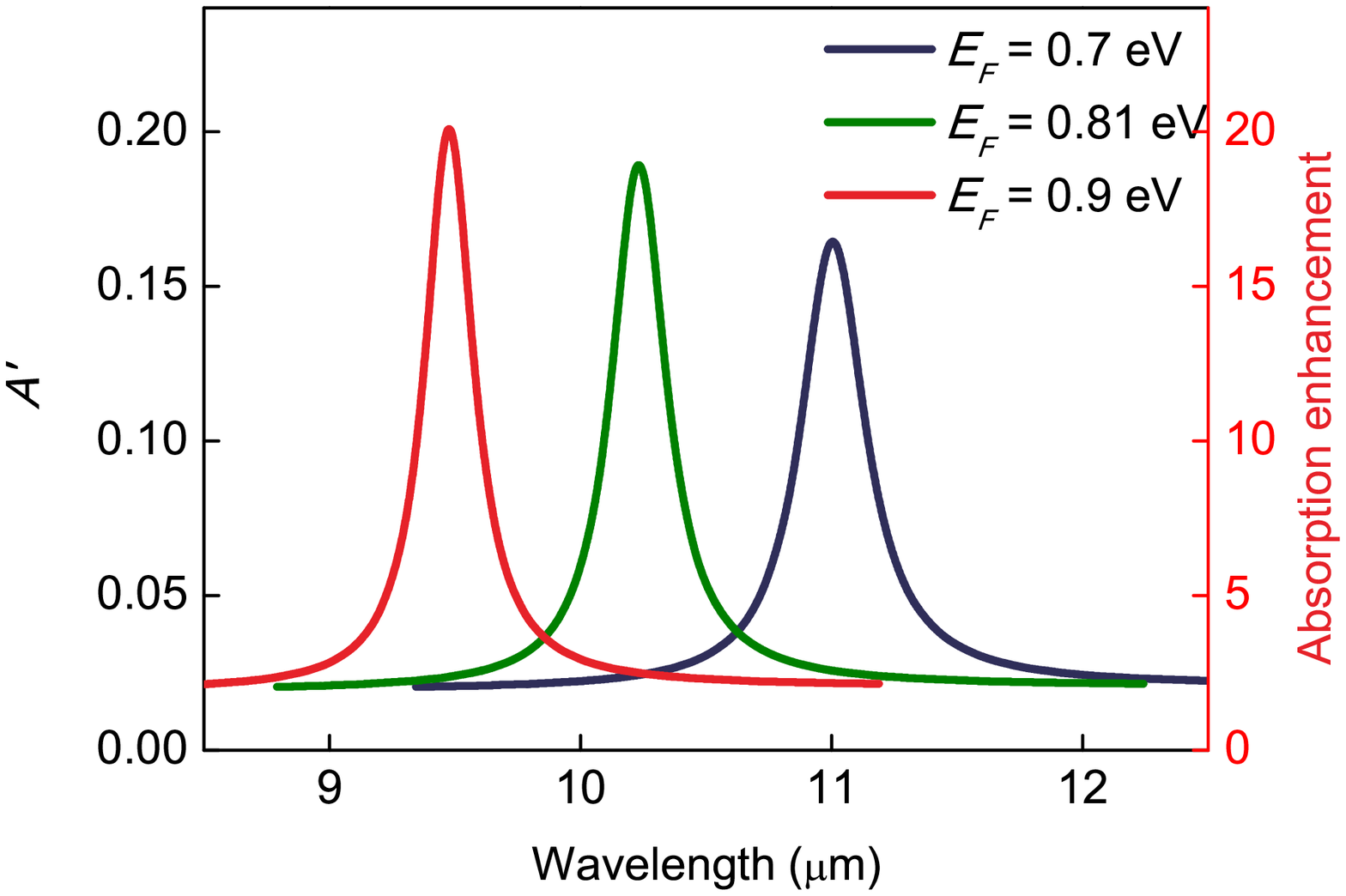}}
\subfigure[]{ \label{fig:6subfig:3} 
\includegraphics[scale=0.3]{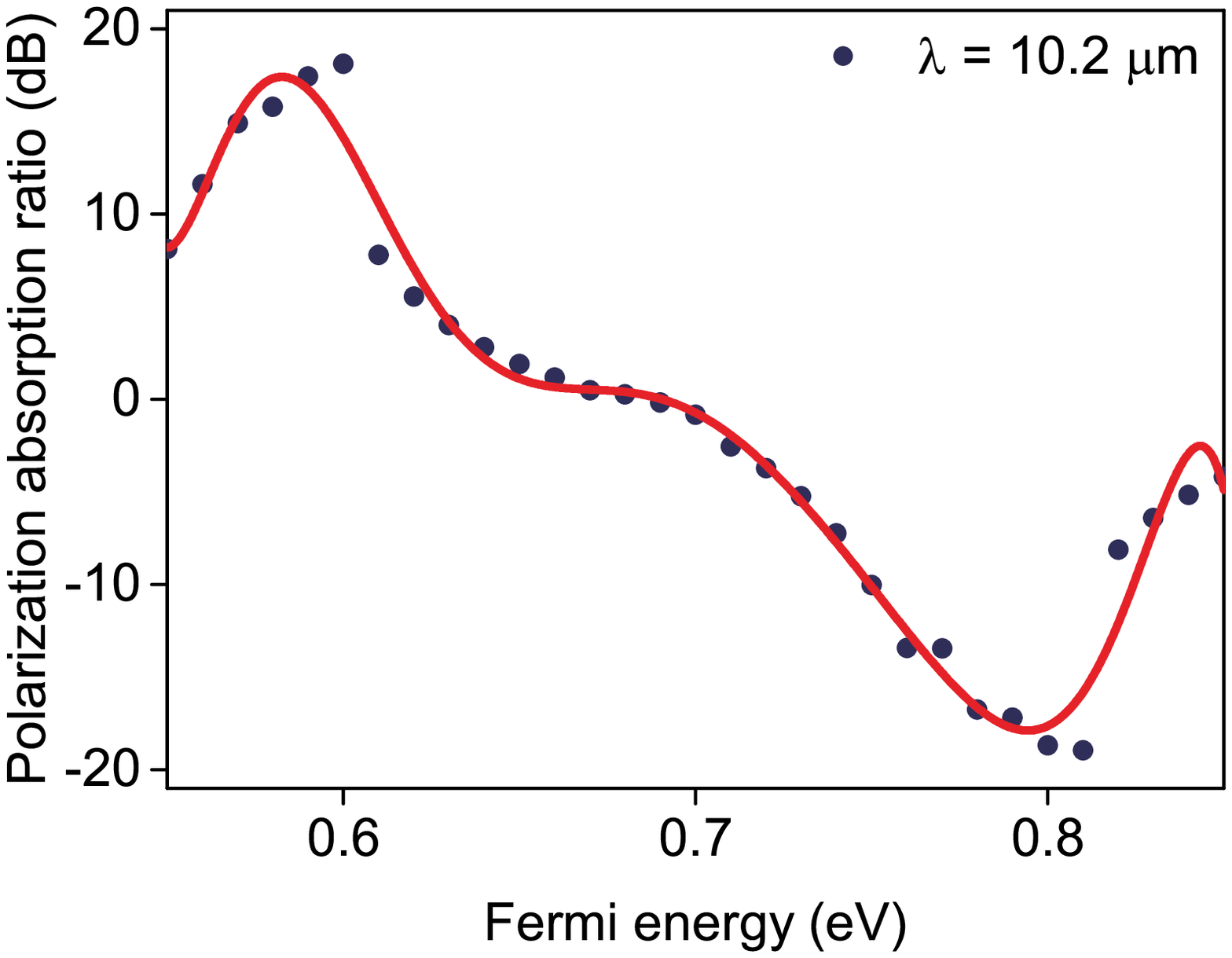}}
\caption{\label{fig:6}The simulated spectra of the absorption in the light-absorbing material $A^{'}$ with different Fermi energies of graphene $E_{F}$ for (a) TM and (b) TE plane wave. The enhancement of the optical absorption in the absorbing layer is also shown. The enhancement factor of the absorption in the light-absorbing material is also shown compared to that in impedance matched media. (c) The dependence of the polarization absorption ratio on the variation of Fermi energy of graphene at the specific wavelength of 10.2 $\micro\meter$.}
\end{figure}

With manipulating the Fermi energy of graphene, the tunable light trapping and absorption enhancement can be realized for these anisotropic graphene-based complementary metamaterials as well. FIG.~\ref{fig:6subfig:1} and \ref{fig:6subfig:2} shows the spectra of absorption in the light-absorbing material for TM and TE plane wave with different Fermi energies of graphene. Similar with the variations in circular nanohole-shaped case, the resonance blue shifts to shorter wavelength and the absorption enhancement increases simultaneously as the Fermi energy increases. Note that an interesting phenomenon can be observed by comparing these two figures: the resonance for TM plane wave with the Fermi energy $E_{F}=0.6$ eV and for TE plane wave with the Fermi energy $E_{F}=0.81$ eV share the exact same resonance wavelength at 10.2 $\micro\meter$, which means either TM or TE plane wave at the specific wavelength can be efficiently absorbed by simply manipulating the Fermi energy of graphene. Here we introduce the polarization absorption ratio defined by $20*\log(A^{'}_{\textrm{TM}}/A^{'}_{\textrm{TE}})$, where $A^{'}_{\textrm{TM}}$ and $A^{'}_{\textrm{TE}}$ are the absorption in the light-absorbing material for TM and TE plane wave, respectively. The dependence of the polarization absorption ratio on the Fermi energy of graphene is explored at the specific wavelength of 10.2 $\micro\meter$. As displayed in FIG.~\ref{fig:6subfig:3}, the polarization absorption ratio shows continuous variations as the Fermi energy of graphene increases. In particular, the ratios with the Fermi energy $E_{F}=0.6$ eV and $E_{F}=0.81$ eV are calculated as 18.1 dB and -19.0 dB. Therefore, the polarization-sensitive absorption of the incident light at the specific wavelength can be dynamically modulated with manipulating the Fermi energy of graphene, which can find potential applications in the next generation of photodetectors with tunable spectral and polarization selectivity in the mid-infrared and THz regimes.

\section{Conclusions}\label{sec4}
In conclusions, we numerically investigate the tunable light trapping and absorption enhancement in graphene-based complementary metamaterials. The excitation of SPR in the monolayer graphene perforated with a periodic array of nanoholes traps the incident light in the near field and leads to absorption enhancement in the light-absorbing material by more than one order of magnitude. Furthermore, polarization-sensitive absorption enhancement can be realized by transforming the shape of complementary resonators from circular nanohole to elliptical nanohole. The tunability of graphene makes it possible to dynamically modulate the absorption enhancement in the light-absorbing material over a broad spectral regime. In particular, either TM or TE plane wave at the specific wavelength can be efficiently absorbed by simply manipulating the Fermi energy of graphene, which is promising for potential applications in the mid-infrared and THz photodetection with spectral and polarization selectivity. Although SPR in the graphene-based metamaterials has mainly been observed in the mid-infrared and THz regimes, it has recently been experimentally demonstrated at much shorter wavelengths ($\sim2$ $\micro\meter$),\cite{wang2016experimental} therefore our proposed hybrid device together with its design principle can be also applied to the near-IR regime.

\begin{acknowledgments}
The author Shuyuan Xiao (SYXIAO) expresses his deepest gratitude to his Ph.D. advisor Tao Wang for providing guidance during this project. SYXIAO would also like to thank Prof. Jianfa Zhang (National University of Defense Technology) for his guidance to the modeling of the light-absorbing semiconductor. This work is supported by the National Natural Science Foundation of China (Grant No. 61376055, 61006045 and 11647122), the Fundamental Research Funds for the Central Universities (HUST: 2016YXMS024) and the Project of Hubei Provincial Department of Education (Grant No. B2016178).
\end{acknowledgments}

\nocite{*}
\bibliography{}

\providecommand{\noopsort}[1]{}\providecommand{\singleletter}[1]{#1}%

\end{document}